# Report on the Aachen OCL Meeting

Achim D. Brucker, Dan Chiorean, Tony Clark, Birgit Demuth, Martin Gogolla,
Dimitri Plotnikov, Bernhard Rumpe, Edward D. Willink, and Burkhart Wolff

**Abstract.** As a continuation of the OCL workshop during the MODELS 2013 conference in October 2013, a number of OCL experts decided to meet in November 2013 in Aachen for two days to discuss possible short term improvements of OCL for an upcoming OMG meeting and to envision possible future long-term developments of the language. This paper is a sort of "minutes of the meeting" and intended to quickly inform the OCL community about the discussion topics.

## 1 Introduction

The meeting started with a short explanation of the OMG standardization process and the plan for the upcoming versions of the OCL standard. Participants agreed to discuss possible short term improvements of OCL for the OMG meeting and to envision possible future long-term developments of the language.

In particular, the participants agreed that
- the Request for Proposals (RFP) for OCL 2.5 will contain only back-ward compatible changes, improvements and clarifications. In particular, there will be no incompatible changes to the concrete syntax of OCL.
- The participants will contribute to a combined submission to the RFP avoiding delays from conflicting submissions.
- Identified problems and possible solutions that involve major changes in OCL will be discussed in an upcoming OCL manifesto that, in its spirit, will be a successor of the "Amsterdam Manifesto on OCL."

In the following, a selection of the topics which were discussed at the meeting are presented briefly. This presentation groups the topics into four main areas:
1. *Core and Execution Semantics* (Sec. 2): this area addresses updates and new features to the core of OCL as well as the question whether OCL should be an executable language or not.
2. *Concrete Syntax:* (Sec. 3): this area discusses extensions of the concrete syntax of OCL that either provide an alternative to already existing OCL expressions or provide a concrete syntax for features that were already available implicitly.
3. *Library Extension* (Sec. 4): this area discusses extensions to the OCL library, e.g., adding new features to datatypes.
4. *OCL Specification Exposition* (Sec. 5): this area comprises general improvements of the OCL specifications.

During the meeting an agreement was not reached for all topics.

## 2 Core and Execution Semantics

In this section, we briefly summarize the topics that were discussed with respect to possible changes to the core of OCL such as support for reflection or the clarification of overloading and method dispatching.

### 2.1 Should OCL be Executable?

Depending on the different use of OCL, the need to ensure that a specification is (efficiently) executable might arise. For example, simulators for OCL or run-time monitoring of OCL constraints require the executability while abstract specifications can often be expressed more concise and elegant using non-executable language constructs. Thus, it was discussed whether OCL as such should be executable or not. Overall, there was a general agreement that adding non-executable constructs is fine, as long as a well-defined and syntactically identifiable subset can be defined that is executable.

### 2.2 Domain of `_.allInstances()`

For `C.allInstances()` there are, in principle, two different interpretation possible that impact the executability of OCL:
- one that returns all actually constructed instances of class `C` that are available in the current system state and
- one that returns all potentially possible instances of the class `C`.

Only the first one interpretation is executable. Moreover, the second interpretation will require infinite sets.

It needs to be discussed, which interpretation should be the standard one and if a second `allInstances()`-like operation should be added to OCL to allow to support both variants.

### 2.3 Reflection

On the one hand, having reflection in a specification language is sometimes useful, in particular for tool builders. On the other hand, reflections causes a lot of challenges, e.g., for the static typing of a language.

During the meeting, it was heavily discussed if (limited) reflection capabilities can be added to OCL without loosing the benefits of a statically typed language. No conclusion was reached here and it remains to be investigated if limited reflections capabilities (e.g., in terms of an extension of the OCL library) or at least being able to query certain properties of the meta-level, should be supported in a future version of OCL.

### 2.4 UML/OCL Data Model and Type Casts

In object-oriented data models with sub-typing, an object has two types at runtime: the static type and the dynamic type. The *static* type (also called *apparent* type) is statically derivable. The *dynamic type* (also called *actual* type) is the real type of an object. With respect to notation, the group prefers *actual type* and *apparent type* (which are also use by the Java community) over static type and dynamic type.

Moreover, it was agreed that casts should be side-effect free and that, in particular, casting an object "up and down" should result in the same object. Note that this is not as obvious as it might sound, e.g., in languages like C++ casts from integers to double are legal but might loose information. Whether casting a value "up and down" should also result in the same value was discusssed with no final conclusion; the discussion mentioned casts between real and integer numbers.

Thus, type casts are only valid between conformant types. Moreover, downcasts might result in `invalid`.

### 2.5 Path Expressions

Many participants of the meeting shared the experience that the "typical OCL user" is not aware of the implicit collect and that the current semantics of path (or navigation) expressions has a lot of subtle corner cases such as navigating over a null valued association end.

As a possible improvement, the introduction of a "dot-question-mark" navigation operator (`_.?_`) was discussed. While the unsafe navigation (`_._`) might contain `null` values (and, thus, in a larger context result in `invalid`), the safe navigation (`_.?_`) filters null values. Adding the safe navigation expression to OCL does not change the language itself, as it is merely a shorthand for a filter expression.

Finally, the need for nested collection was discussed and it was agreed that this needs to be elaborated in the upcoming OCL manifesto. Moreover, it was agreed that the default behavior should be the automatic flattening during a navigation.

### 2.6 Template Types (Generics)

UML supports templates type and OCL at least specifies the collection types using templates. Still, there is no possibility for user-defined functions or classes that use template types. Thus, the upcoming OCL manifesto needs discuss in more detail the semantics of templates (e.g., similar to C++ or Generics in Java) in general and in particular the casting and sub-typing relations between template types.

### 2.7 Recursive Definitions

The handling of recursive definitions in general and recursive functions in particular is not clear. Up to now, the OCL standard allows for defining recursive functions by re-using the function name in the body or post-condition. Moreover, the standard requires that the recursion terminates.

The group discussed means for defining a measurement that allows for easily determining if a recursion terminates (i.e., is a well-defined recursion as well as the introduction of a new keyword (e.g., `letrec`) to make (mutual) recursive definitions explicit.

### 2.8 Overloading and Dispatch

Currently, the OCL standard does not define the dispatch of overloaded methods. Thus, both the static and dynamic dispatch needs to be defined. For the *static dispatch* the obvious approach is to take the highest common superclass and make the single dispatch on it. The *dynamic dispatch* requires a more in-depth analysis to keep, e.g., the collection types as precise as possible (and not just collection of `OclAny`). To avoid unintential specification of to broad collection types, the need for explicit type intention might be necessary (e.g. `Set{}::Set(Integer)`).

### 2.9 Non-Determinism in Specifications

It was also discussed that, in general, the evaluation of an OCL specification is non deterministic, e.g., evaluating an expression containing `_->any(_)` might even for different evaluations in the same tool yield a different result. Also functions like `_->asSequence()` are underspecified, i.e., the detailed behavior is defined by the tool implementer.

This non-determinism is desired on the level of the specification. Still, tool vendors might choose to "determinize" an implementation to, e.g., always return the first satisfying element for an any-call on a sequence.

### 2.10 Multiple Inheritance

As UML supports multiple inheritance, OCL has to support multiple inheritance as well. For "closed" OCL specifications, the usual problems with multiple inheritance should not be prevalent, as there is no method implementation. Still, the detailed consequences and possible problems need to be investigated further. It became already clear that multiple inheritance impacts the "open world" assumption and, thus, impacts the re-usability of verification results in case of extending a specification using inheritance.

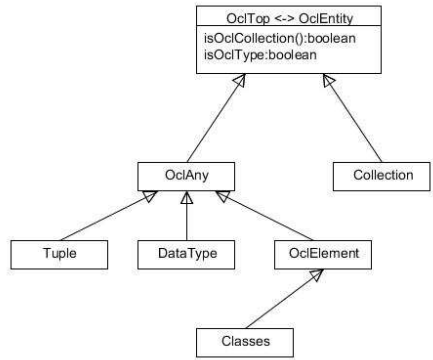

**Fig. 1.** Proposal for the new common supertype

### 2.11 Exception Elements or Multi-valueness of the Logic

Since its beginning, OCL is based on a three-valued logic. In fact, all datatypes (including `Boolean`) include an exception element (called `OclUndefined` in earlier versions of the standard). Recent versions of the OCL standard added an additional exception element, requiring a four-valued logic. It seemed that in the current documents, the description of the conceptual difference between both exceptional value and its practical employment could be improved.

As multi-valued logics, compared to traditional two valued logics, required more sophisticated tools, the need for a multi-valued logic was discussed in detail. At the end, a relative majority was for staying with the four-valued logic in a discussion whether to use two, three, or four truth values.

### 2.12 `OclAny` Conformance and `OclEntity`

The conformance to `OclAny` in general and in particular that the collection types conform to `OclAny` was discussed as well. One reason that this conformance creates problems is the fact, that the interface of `OclAny` is too rich. Since a common super type is required for the polymorphic behavior, the introduction of a new super class with only two methods `isOclCollection()` and `isOclType()` was discussed (cf. Fig. 1).

### 2.13 Framing

Traditionally, OCL operation contracts do only specify the intended changes to the system state. In general, there is no guarantee that other parts of the system remain unchanged. In particular, the default post condition `true` allows arbitrary changes to the system state.

To solve this problem, the introduction of a new method `_->modifiesOnly()` was discussed. This methods should allow to explicitly specify (if necessary, using a recursive predicate that actually computes) the set of objects that are allowed to be changed.

### 2.14 Bounded Types

The incompatibility between UML support for bounded types such as for example `String[0..5] {unique}` and OCL support for collection types was discussed. It was felt that simple support for cardinality as a collection type annotation was adequate and that the restrictions imposed by these bounds would be erased as soon as a bounded type participates in an evaluation.

## 3 Concrete Syntax

In this section, we briefly summarize the topics that were discussed with respect to possible extensions of the concrete syntax.

### 3.1 Type Construction

While OCL has a nice notation for constructing collections (e.g., `Set{42,37}`), it is not possible to construct arbitrary objects in a similar way.

As OCL expression should, in general, be side-effect free. Thus, it was discussed that only a limited form of object constructors should be supported. In more detail, an approach based on tuples (also called records) was discussed. The core idea is to support the construction of a record-based representation of objects, i.e., without an object id that actually would change the system state, shall be supported. While this is only a limited way of construction objects, there was a general agreement that this should provide a useful solutions for the most important use cases. Still, the details need to be investigated.

### 3.2 Lambda Types and Expressions

At some places, e.g., iterator expressions, OCL already supports $\lambda$-expressions— or at least something that resembles $\lambda$-expressions very closely. In general, there was an agreement that a future version of OCL should support $\lambda$-expressions as first class citizen. For this, some details of the semantics as well as the concrete syntax needs to be clarified.

### 3.3 Pattern Matching

Pattern matching has to proven a very useful concept in many languages. It was discussed to which extent the OCL syntax can be extended, without breaking backward compatibility, to support pattern matches. The details need to be clarified and discussed in the upcoming OCL manifesto.

### 3.4 Collection Comprehension

There was a general agreement that a (mathematical) set comprehension-like notation would be useful. On the long term, this should be discussed in the upcoming OCL manifesto and included in a future version of OCL. Finally, collection comprehension provides, naturally, a concise syntax alternative for `_->select(_)`, `_->reject(_)`, and `_->collect(_)`.

### 3.5 Variable Arguments for Operations

Support for operations with variable argument lists (e.g., similar to `printf(...)` in C) was discussed. Overall, this is only syntactic sugar and can be supported easily.

### 3.6 Java-like Syntax of OCL

Finally, complete alternative syntax variants such as a Java-like syntax or a mathematical OCL were discussed. While such syntax variants are desirable in specific use cases (e.g, when using OCL in the context of Java, a Java-like syntax will most likely increase the adoption of OCL by Java users), there was an agreement that OCL, as a generic object-oriented constraint language, needs to provide a concrete syntax that is not strongly linked to a specific use case. Thus, tool vendors are free to provide additional alternative syntax variants while the OCL standard will, most likely, concentrate on the existing concrete syntax (and extensions thereof).

## 4 Library Extension

In this section, we briefly summarize the topics that were discussed with respect to possible extensions of the OCL library.

### 4.1 Regular Expressions

Regular expressions have proven to be useful in many languages. During the meeting the extension of the type `String` with operations for construction and matching regular expressions was discussed.

### 4.2 Implementation-level Datatypes

OCL types are specification-level types that are not restricted by the limitations of a concrete implementation. For example, the OCL type `Integer` represents the "mathematical" Integers, i.e., unbounded Integers. In contrast, programming languages usually prefer fixed-sized machine-arithmetic (e.g., Integers based on a 32 bit two's-complement representation).

As the semantics of implementation-level types differs significantly from the semantics of the specification-level types, it was suggest to add machine representations to the OCL library. This would allow to distinguish specification-level and implementation-level UML/OCL specifications as well as to analyze the conformance relations between the different abstraction level.

### 4.3 Tuple Join

As many people use OCL to formulate queries over UML data models, it seems natural to request SQL-like features such as joining of tuples. After a longer discussion, it was concluded that the static type system of OCL (including the need for supporting subtyping) makes it difficult to support tuple joins similarly to OCL. The details need to be investigated further.

## 5 OCL Specification Exposition

Finally, general improvements of the OCL specification were discussed. This included the need for updating the motivational examples of the standard (e.g., to also include the use of OCL in the context of state machines). This area comprises general improvements of the OCL specifications or best practices in writing OCL specifications.

### 5.1 Updating the Semantics Part of the Standard

Currently, the OCL standard contains two chapters that are related to the semantics of the language:
– Chapter 10: "Semantics Described Using UML"
– Annex A: "Semantics"

After a longer discussion, it was concluded that the most important part of chapter 10 are the well-formedness rules for OCL expression. It was decided that this part should be preserved (most likely integrated into different chapters of the mandatory part in the standard) in upcoming versions of the OCL standard. Additionally, the informative annex should contain an updated formal semantics of the core of OCL. The ultimate goal is to provide a machine checked semantics, e.g., using Isabelle/HOL (this could be based on Featherweight OCL).

### 5.2 Annotated EBNF Grammar

It was decided that the next version of the OCL standard should include a machine readable EBNF (e.g, using an annotated grammar similar to XText, MontiCore or EMFText) of the concrete syntax of OCL.

### 5.3 Ghost Fields

Ghost fields, e.g., attributes that are only available for modeling purposes and not mapped to an implementation, have to be proven useful in languages such as JML or Spec#. Thus, it was suggested to extend OCL with support for ghost fields as well.

After an intense discussion it was recognized that OCL most likely already supports a similar concept using the `def` context declaration. This needs to be investigated further and described in more detail in the upcoming OCL manifesto.

### 5.4 Inclusion of Specification Patterns in the OCL Standard

It was generally acknowledged that the motivational examples in the standard should be improved. More particular, it was suggest that the standard should also include a section explaining best practices and misuse patterns of OCL, i.e., guidelines on how to write good OCL specifications.

## 6 Conclusion

The two-day meeting in Aachen provided a platform for a large number of fruitful discussions around OCL in general and both short-term and long-term opportunities for improving OCL in particular.

Overall, the discussions showed that OCL is a matured language that fills a gap in the landscape of specialized (object-oriented) constraint or specification languages such as JML (for Java) or Spec# (for C#). Even though OCL is nearly 20 years old, there are still many open issues ranging from fixing bugs in the standard to long-term research challenges. Consequently, follow-up discussions and meetings are expected and the participants plan to provide a more detailed analysis and recommendation in form of a successor of the "Amsterdam Manifesto on OCL."